\documentclass[journal=langd5,manuscript=letter]{achemso}
\author{Manish Vashishtha}
\affiliation{Department of Chemical Engineering, 
             Indian Institute of Technology Delhi, New Delhi, India}
\author{Prabhat K. Jaiswal}
\affiliation{School of Physical Sciences, 
             Jawaharlal Nehru University, New Delhi, India}
\author{Rajesh Khanna}
\affiliation{Department of Chemical Engineering, 
             Indian Institute of Technology Delhi, New Delhi, India}
\email{rajkh@chemical.iitd.ac.in}
\author{Sanjay Puri}
\affiliation{School of Physical Sciences, 
             Jawaharlal Nehru University, New Delhi, India}
\author{Ashutosh Sharma}
\affiliation{Department of Chemical Engineering, 
             Indian Institute of Technology Kanpur, Kanpur, India}

\title{Spinodal Phase Separation in Liquid Films with Quenched Disorder}
\begin{document}
%%%%%%%%%%%%%%%%%%%%%%%%%%%%%%%%%%%%%%%%%%%%%%%%%%%%%%%%%%%%%%%%%%%%%%%
\begin{abstract}
We study spinodal phase separation in unstable thin liquid films on
chemically disordered substrates via simulations of the thin-film equation.
The disorder is characterized by immobile patches of varying size and 
Hamaker constant. The effect of disorder is pronounced in 
the early stages (amplification of fluctuations), remains during the 
intermediate stages and vanishes in the late stages (domain growth). 
These findings are in contrast to the well-known effects of quenched 
disorder in usual phase-separation processes, viz., the early stages 
remain undisturbed and domain growth is slowed down in the asymptotic 
regime. We also address the inverse problem of estimating disorder by 
thin-film experiments.
\end{abstract}
%%%%%%%%%%%%%%%%%%%%%%%%%%%%%%%%%%%%%%%%%%%%%%%%%%%%%%%%%%%%%%%%%%%%%%%
\section{Introduction}
Consider a system which is rendered thermodynamically unstable by a sudden 
change of parameters, e.g., temperature, pressure. The subsequent evolution 
of the system is characterized by the emergence and growth of domains 
enriched in the preferred phases 
\cite{pw09}. 
This domain growth process is of great importance in science and technology. 
In this context, an important paradigm is the phase-separation kinetics of 
an initially homogeneous binary (AB) mixture. If the initial concentration 
fluctuations grow spontaneously, the evolution is referred to as 
{\it spinodal decomposition}. The segregating mixture evolves into coexisting 
domains of A-rich and B-rich phases. In pure (disorder-free) systems, these 
domains are characterized by an increasing length-scale, $L(t)$, which grows 
as a power-law, $L(t)\sim t^\phi$ in time $t$. The exponent $\phi$ depends on 
the transport mechanism. For diffusive transport, $L(t)\sim t^{1/3}$, which 
is known as the Lifshitz-Slyozov (LS) growth law and is based on the 
{\it evaporation-condensation} mechanism
\cite{ls61}. 

\begin{figure}[ht]
\begin{center}
\includegraphics*[width=0.8\textwidth]{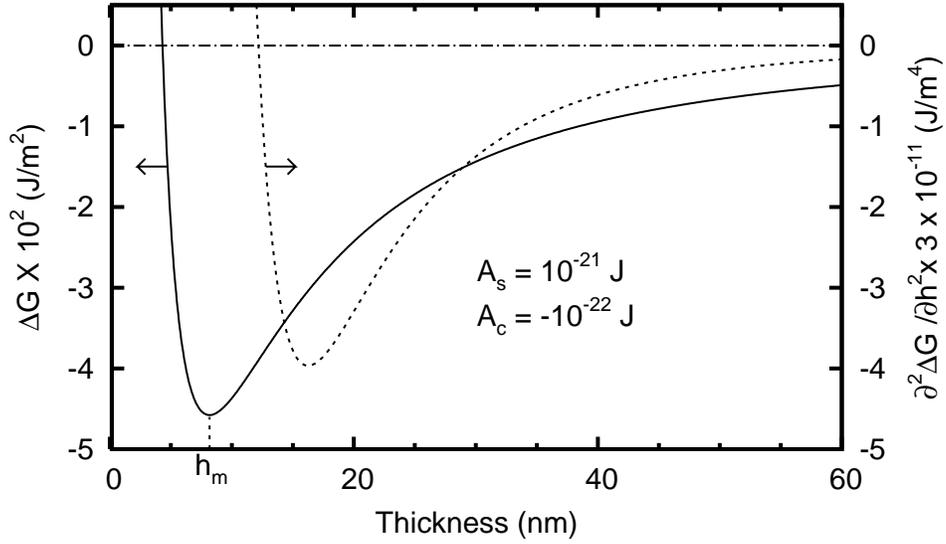}
\end{center}
\caption{Variation of the free energy per unit area 
         ($\Delta G$, solid line) and spinodal parameter 
         ($\partial^{2} \Delta G/ \partial h^{2}$, dashed line) 
         with film thickness. The parameter values are 
         $R = -0.1$ and $\delta=10$ nm.
         Spinodal phase separation takes place for $h > 12.5$ nm, where 
         $\partial^{2} \Delta G/ \partial h^{2} < 0$. 
         }
\label{fig1}
\end{figure}

In recent work 
\cite{pre10}, 
we have emphasized the analogies and differences between usual phase-separation 
kinetics and the {\it spinodal phase separation} (SPS) dynamics of an unstable 
thin liquid film ($< 100$ nm) on a pure substrate. Typically, random fluctuations 
in the free surface of initially flat films grow and evolve into distinct 
morphological phases, viz., a thinner low-curvature flat film phase, and a 
thicker high-curvature droplet phase. Spinodal growth occurs when the excess 
intermolecular free energy $\Delta G(h)$ shows a minimum and the spinodal parameter, 
$\partial^{2} \Delta G/ \partial h^{2}|_{h = {h_0}} < 0$. Here, $h$ is the film 
thickness, and $h_0$ is the average thickness 
\cite{as93, as98}. 
In~{\ref{fig1}}, 
we show a typical form of $\Delta G$ vs. $h$ and the corresponding spinodal 
parameter, $\partial^{2} \Delta G/ \partial h^{2}$ vs $h$. The double-tangent 
construction for $\Delta G$ in~{\ref{fig1}} shows that the film segregates 
into phases with $h=h_m$ and $h = \infty$. (This should be compared with 
phase-separation problems, which are described by a double-well potential, 
and there are two possible values for the equilibrium composition 
\cite{pw09, ab94}.) 
In an alternative scenario, commonly known as {\it true dewetting}, the 
thin film breaks up into dewetted dry spots (with $h=0$) surrounded by 
undulating liquid ridges and there is no SPS
\cite{as93}.
Our recent work 
\cite{pre10} 
on SPS on homogeneous substrates has shown that the number density of local 
maxima in the film's surface (defects), $N_d$, is a suitable marker to identify 
the early, late and intermediate stages of SPS in thin films. We have also 
established that the kinetics of the early stages can be described by a universal 
evolution of this marker with a suitably rescaled time, and the late stages can 
be described by the LS growth law. 

The above discussion has focused on SPS in pure systems with homogeneous 
substrates. However, real experimental systems never satisfy this condition.
The substrates contain physical and chemical impurities (disorder), which 
can be immobile (quenched) or mobile (annealed). We have some understanding of 
the alternative scenario of true dewetting on disordered substrates 
\cite{kk2000},
but the problem of SPS with disorder is yet to be addressed. This letter 
investigates SPS in unstable thin liquid films on substrates with quenched 
chemical disorder, and highlights novel features which arise due to the presence 
of disorder. The inverse problem of estimating the disorder through thin-film 
experiments is also addressed, as this would be very useful for experimentalists.
%%%%%%%%%%%%%%%%%%%%%%%%%%%%%%%%%%%%%%%%%%%%%%%%%%%%%%%%%%%%%%%%%%%%%%%
\section{Model and Simulation}
The starting point of our study is the thin-film equation which models the 
evolution of the film's surface $h(\vec{x},t)$ in supported thin liquid films. 
This is derived by simplifying the equations of motion via the 
{\it lubrication approximation} 
{\cite{ruck74}}. 
The resulting equation is analogous to the Cahn-Hilliard (CH) equation 
\cite{ch58} 
of phase separation with an $h$-dependent mobility 
{\cite{vsm93}}, 
$M(h) = h^{3}/(3\mu)$ (corresponding to Stokes flow with no slip). 
The total free energy is 
$F_s[h] = \int [\Delta G(h) + \gamma (\vec{\nabla} h)^2/2] d\vec{x} \equiv F_e + F_i $, 
where $F_{e}$ is the overall excess free energy and $F_{i}$ is the
interfacial free energy. The quantities $\gamma$ and $\mu$ refer to surface 
tension and viscosity of the liquid film, respectively. The corresponding 
CH equation is
\begin{equation}
\frac{\partial}{\partial t} h(\vec{x},t)
 = \vec{\nabla} \cdot \left[M\vec{\nabla}\left(\frac{\delta F_s}{\delta h}\right)\right] 
 = \vec{\nabla} \cdot 
   \left[\frac{h^{3}}{3 \mu} \vec{\nabla}
    \left( \frac{\partial \Delta G}{\partial h} - \gamma \nabla^{2}h
    \right)
   \right], 
\label{dtfe}
\end{equation}
where all gradients are taken in the plane of the substrate. The potential 
$\Delta G(h)$ usually combines a long-range attraction and a short-range 
repulsion. The results presented here correspond to a long-range van~der~Waals 
attraction due to the substrate, and a comparatively short-range 
van~der~Waals repulsion provided by a nano-coating on the substrate 
\cite{as96}: 
$\Delta G = -A_{c}/12\pi h^{2} -A_{s}/12\pi(h+\delta)^{2}$. 
This is the potential shown in~{\ref{fig1}}. Here, $A_{s}$ and $A_{c}~(= R A_s)$ 
are the effective Hamaker constants for the system, which consists of the fluid 
above the film, the liquid film, and a solid substrate ($s$) or coating 
material ($c$). The thickness of the nano-coating is $\delta$. 

We can reformulate~\ref{dtfe} in a dimensionless form as follows:
\begin{equation}
\frac{\partial}{\partial T} H(\vec{X},T)= \vec{\nabla} \cdot \left[H^{3} \vec{\nabla}
\left( \frac{2 \pi h^{2}_{0}}{|A_{s}|} \frac{\partial \Delta G}{\partial H} - \nabla^{2}H
\right) \right]. 
\label{tfe}
\end{equation}
In~\ref{tfe}, $H=h/h_0$, where $h_0$ is the mean film thickness; 
$\vec{X}=\vec{x}/\xi$, where 
$\xi=\left( 2 \pi \gamma/ \mid A_{s} \mid\right)^{1/2} h^{2}_0$ 
is the characteristic scale for the van der Waals case; and $T=t/\tau$, 
where $\tau=\left( 12 \pi^{2} \mu \gamma h_0^{5}/ A_{s}^{2}\right)$. 
The excess free-energy term now has the form 
\begin{equation}
 \frac{2 \pi h^{2}_{0}}{\mid A_{s} \mid} 
  \frac{\partial \Delta G}{\partial H} = \frac{1}{3} 
  \left[ \frac{1-R}{\left(H + D\right)^{3}}+ \frac{R}{H^{3}}\right], 
\label{phi}
\end{equation}
where $D = \delta/h_{0}$ is the nondimensional coating thickness.
The linear stability analysis of~\ref{tfe} for fluctuations 
about $H=1$ predicts a dominant spinodal wave of wave-vector $k_M$ and 
time-scale $T_M$ with  
$$ k_{M} = \frac{2\pi}{L_M} = {\sqrt{-\frac{\pi h_0^{2}}{\mid A_{s}
\mid} \frac{\partial^{2}\Delta G}{\partial H^{2}}\bigg|_{H = 1}}}
\equiv \sqrt{\frac{\alpha}{2}} \,\, ,$$
\begin {equation}
 T_M = \frac{4}{\alpha^2}. 
\label{km}
\end{equation}

\begin{figure}[ht]
\begin{center}
\includegraphics*[width=0.8\textwidth]{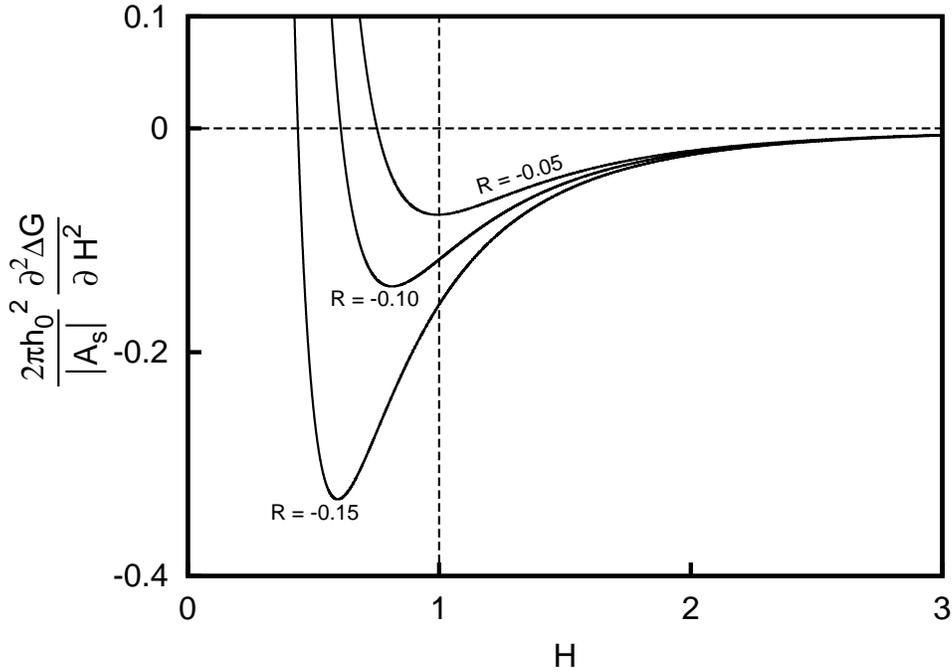}
\end{center}
\caption{Variation of dimensionless spinodal parameter with 
         nondimensional film thickness for three values of 
         $R = -0.15, -0.1$ and $-0.05$.
         }
\label{fig2}
\end{figure}

For a homogeneous substrate, the parameters in~\ref{tfe}-\ref{phi} are spatially 
uniform. We model the quenched disorder via fixed patches of varying size $P$ 
with relative Hamaker constant $R$. The mean values of $P$ and $R$ are $P_m$ 
and $R_m$. These parameters are uniformly distributed in the intervals 
[$P_m-\delta P, P_m+\delta P$] and [$R_m-\delta R, R_m+\delta R$], with 
$\delta P/ P_m = P_d$ and $\delta R/ R_m = R_d$. 
In~{\ref{fig2}}, 
we plot the spinodal parameter 
$(2\pi h_0^2/\mid A_s \mid) \partial^2 \Delta G/ \partial H^2$ vs. $H$ for 
three representative values of $R$. These correspond to the mean value 
$R_{m} = -0.1$, and the extreme deviations for $R_{d} = 0.5$. Notice that 
a small change in $R$ results in a major change of $k_M$ (or $L_M$) and 
$T_M$ in~\ref{km}. 

\begin{figure}[ht]
\begin{center}
\includegraphics*[width=0.8\textwidth]{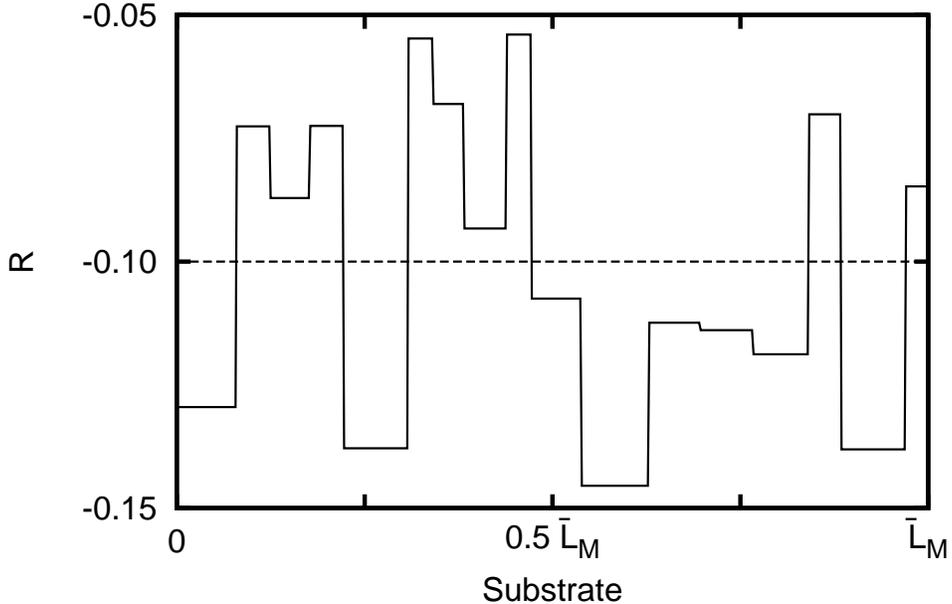}
\end{center}
\caption{Schematic of a chemically disordered substrate. 
         The disorder is quenched and is characterized by 
         $R_{m} = -0.1$ ( horizontal dashed line), $R_{d} = 0.5$,
         $P_{m} = \bar{L}_{M}/16$ and $P_{d} = 0.5$.}
\label{fig3}
\end{figure}

We numerically solve~\ref{tfe} in $d=2,3$ starting with a small-amplitude 
($\simeq 0.01$) random perturbation about the mean film thickness $H = 1$. 
The system size in $d=2$ is $n\bar{L}_{M}$, where $\bar{L}_M$ is the 
dominant wavelength for $R=R_m$ ($n$ ranges from 16 to several thousands). 
The system size in $d=3$ is ${(16 \bar{L}_M)}^2$. We apply periodic 
boundary conditions. A 512-point grid per $\bar{L}_{M}$ was found to be 
sufficient when central-differencing in space (with half-node interpolation) 
was combined with {\it Gear's algorithm} for time-marching, which is 
convenient for stiff equations. The parameters $D$, $R_{m}$ and $R_{d}$ 
were chosen so that the film is spinodally unstable at $H = 1$ 
(i.e., $\alpha > 0$) for all values of $R$. A typical spatial variation of 
$R$ with $R_m=-0.1$ and $R_d=0.5$ is shown in~{\ref{fig3}}. $P_{m}$ was 
taken as $f \bar{L}_{M}$, where $f$ varies from 1/16 to 16. Thus, the 
size of the patches with fixed $R$ varied from being much smaller than 
the spinodal length-scale to being much bigger than it.  
%%%%%%%%%%%%%%%%%%%%%%%%%%%%%%%%%%%%%%%%%%%%%%%%%%%%%%%%%%%%%%%%%%%%%%%
\section{Results and Discussion}
\begin{figure}[ht]
\begin{center}
\includegraphics*[width=0.8\textwidth]{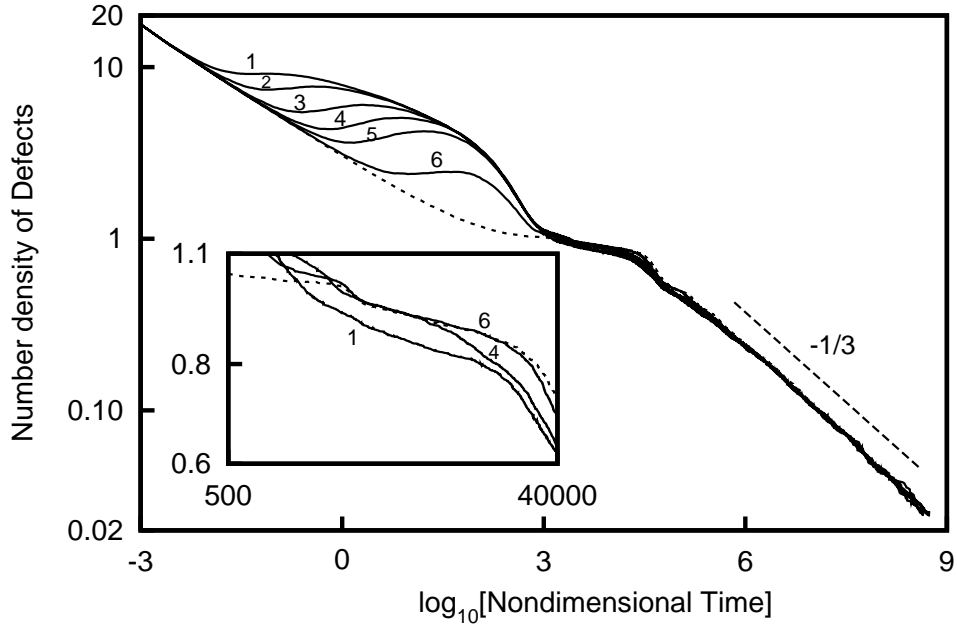}
\end{center}
\caption{Variation of the number density of defects ($N_d$) with nondimensional 
         time ($T$) for a $d=2$ system of size 2048 $\bar{L}_{M}$ and $D = 0.5$. 
         Curves 1 to 6 present results for $R_m=-0.1$ and 
         $R_{d} =0.50,\,0. 25,\, 0.10,\,0.05,\, 0.03$ and $0.01$, 
         respectively. The LS $-1/3$ slope for late stages is shown by
         the dashed line.
         The dotted curve corresponds to the case with zero disorder (base case). 
         The inset shows the magnified intermediate stage for
         $R_{d} =0.50,\,0.05$ and $0.01$. The other parameter values are 
         $P_{m}=\bar{L}_{M}/16$, $P_{d}=0.5$.
         }
\label{fig4}
\end{figure}

\begin{figure}[htb]
\begin{center}
    \begin{tabular}{cc}
      \resizebox{80mm}{!}{\includegraphics{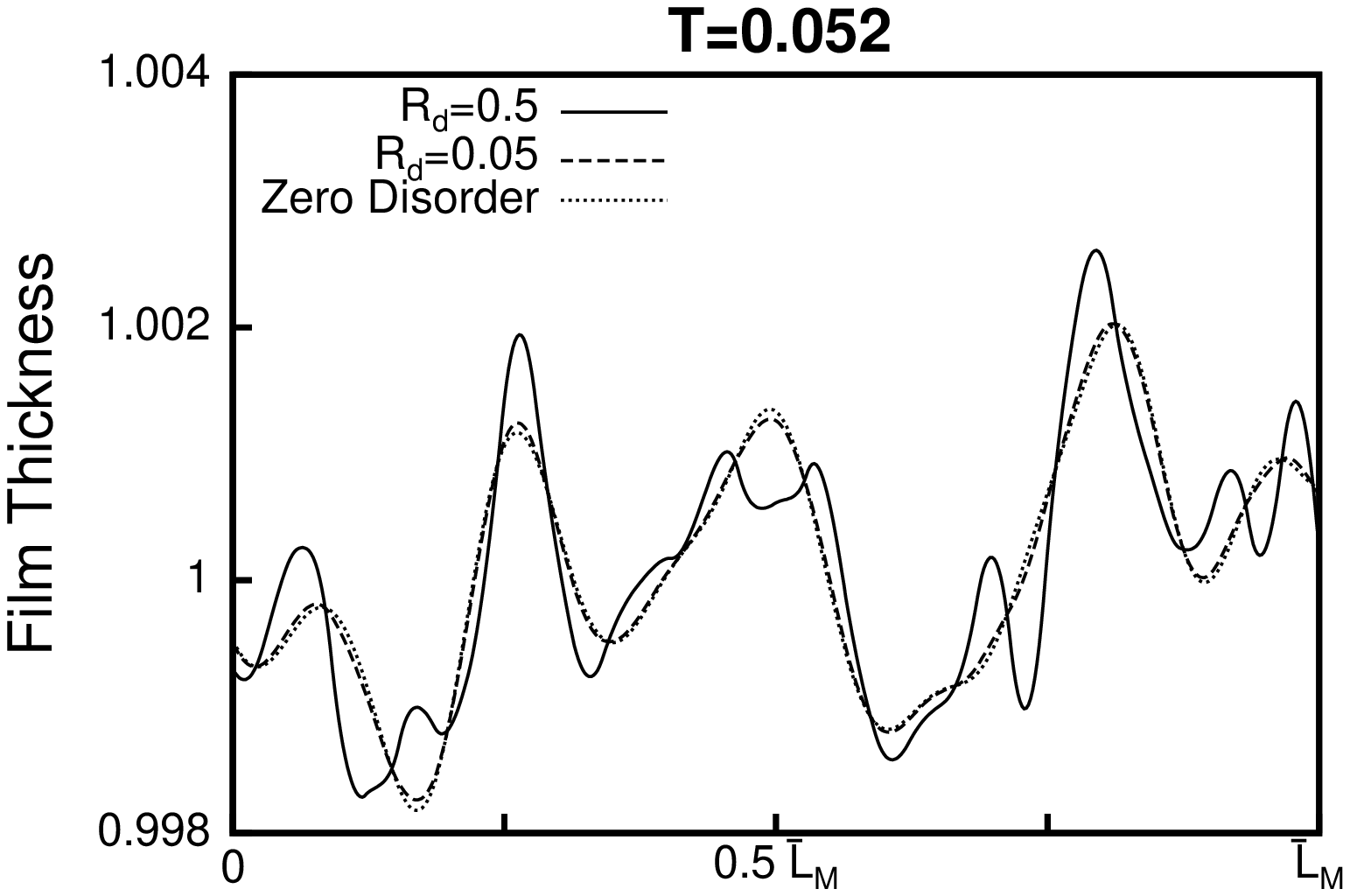}} &
      \resizebox{80mm}{!}{\includegraphics{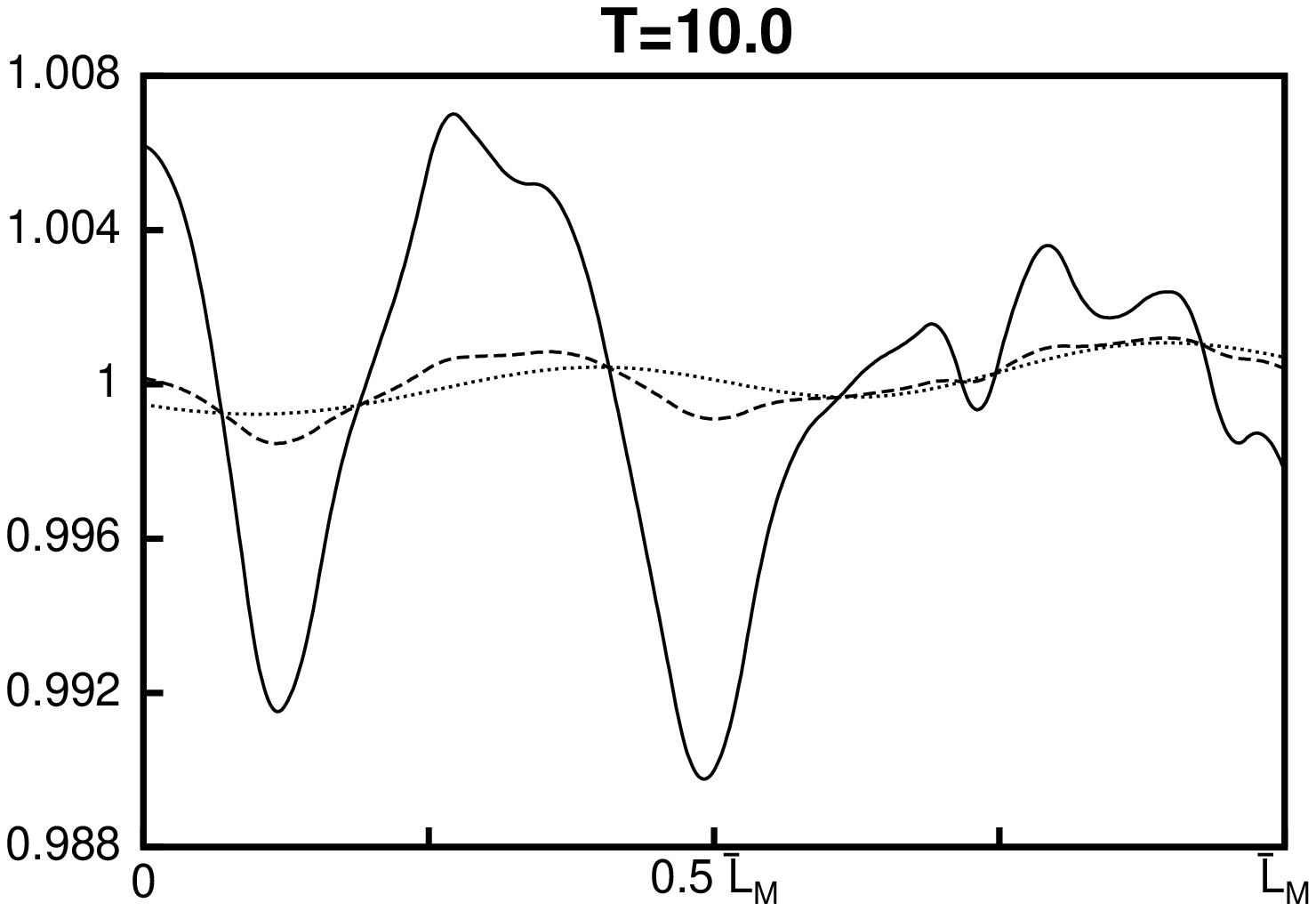}} \\
      \resizebox{80mm}{!}{\includegraphics{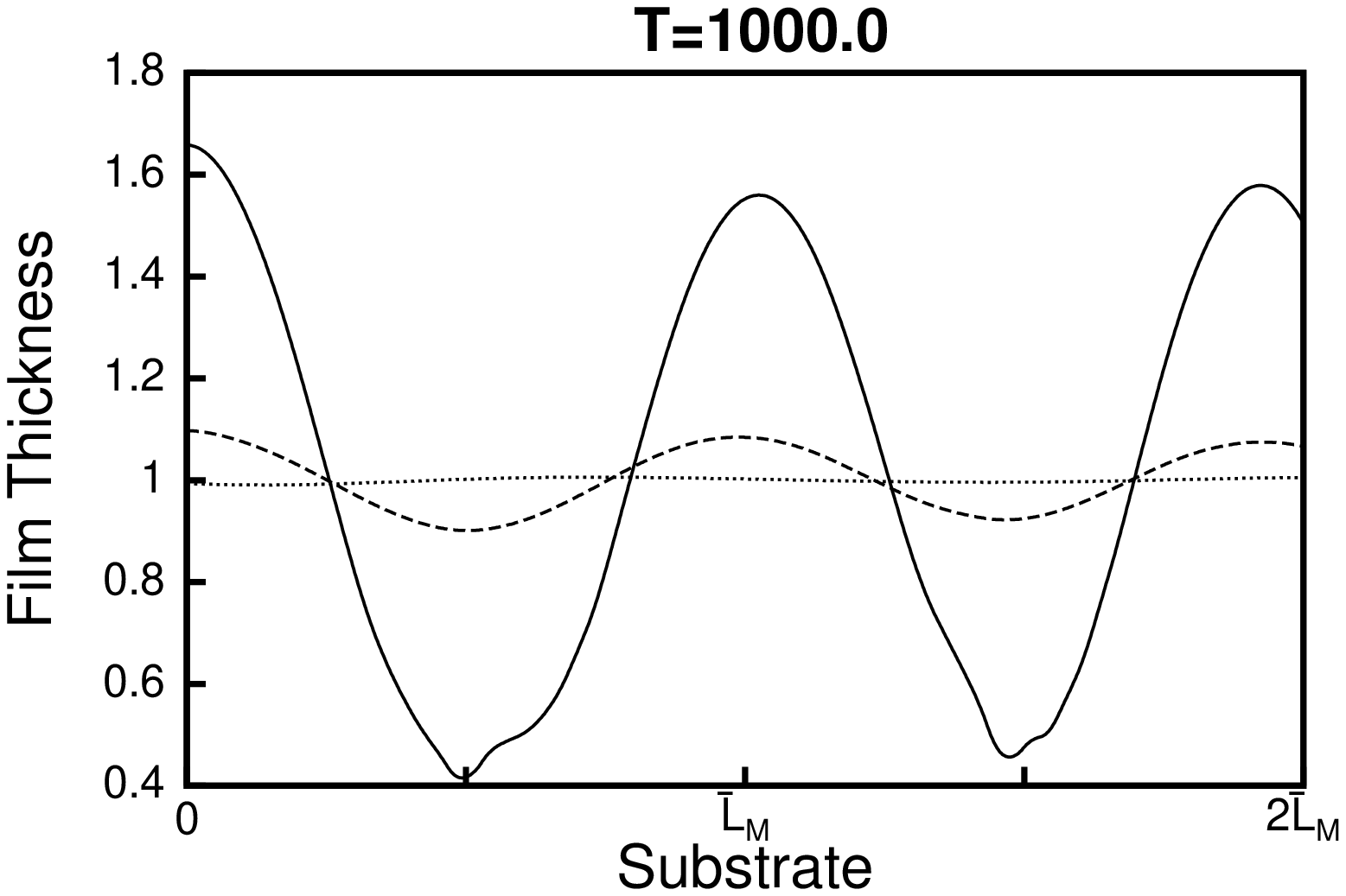}} &
      \resizebox{80mm}{!}{\includegraphics{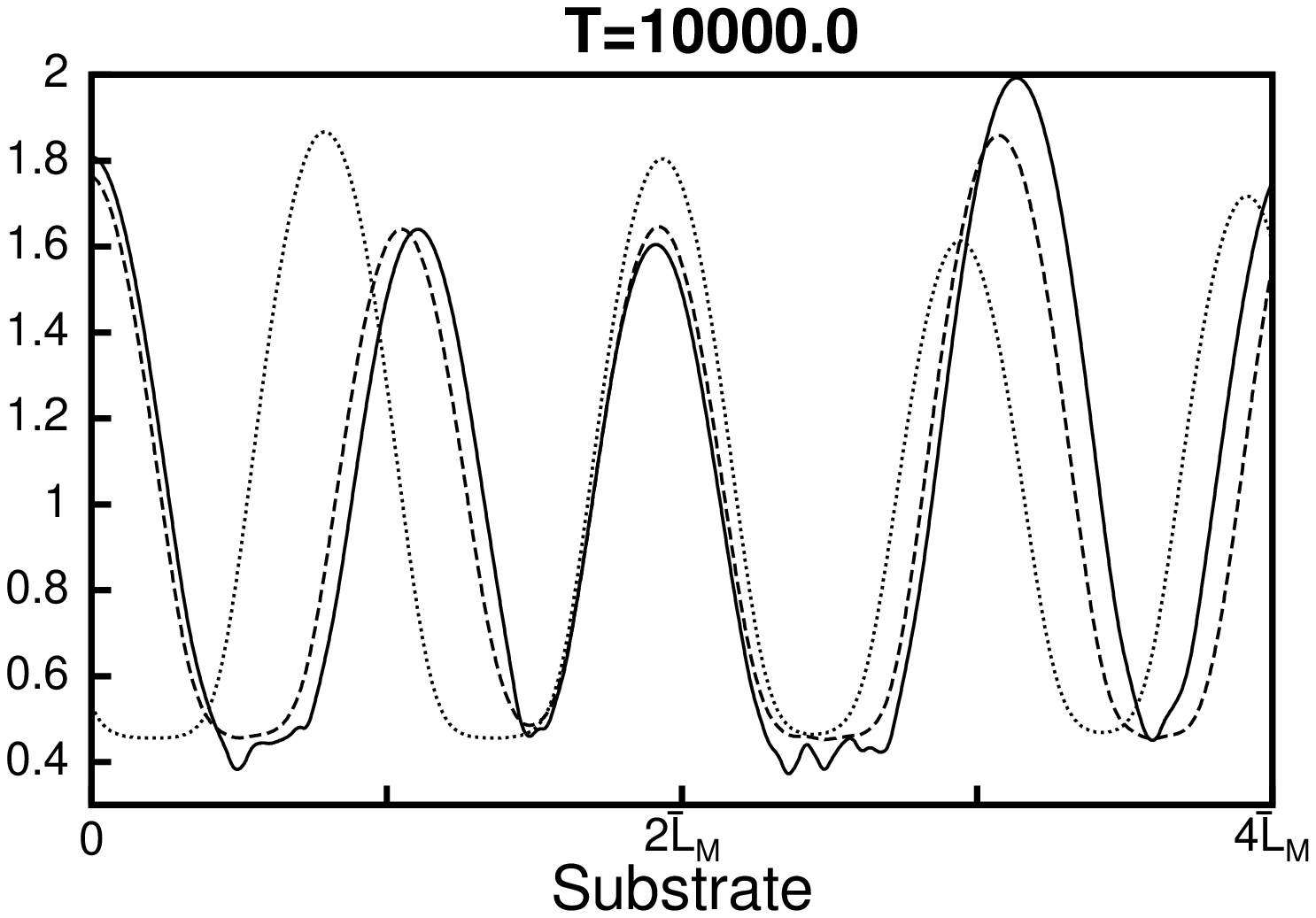}} \\
    \end{tabular}
    \caption{Morphological evolution of the film at different
             nondimensional times, as specified. We show results 
             for $R=-0.1$ and $R_d=0.5, 0.05$ and the pure case. 
             The other parameter values are the same as 
             in~{\ref{fig4}}.}
    \label{fig5}
  \end{center}
\end{figure}

All the three markers of the process, viz., number density of defects 
($N_{d}$), morphology and energy are affected by the strength ($R_{d}$) 
as well as the length-scale ($P_{m}$) of the chemical disorder. Notice 
that $N_d^{-1}$ characterizes the typical domain size. In~{\ref{fig4}}, 
we plot $N_d$ vs. $T$ on a log-log scale. The effect of disorder is 
pronounced in the early stages, persists during the intermediate stages, 
and vanishes in the late stages. The evolution of $N_{d}$ briefly 
follows the results of the pure case ($R_d=0$), but is seen to diverge 
while still in the early stages.
However, surprisingly, the plot of $N_d$ vs. $T$ reverts to the pure case in 
the late stages. This is the central result of this letter, and will be 
discussed shortly. The deviation in the intermediate stages is seen in the 
magnified view in~{\ref{fig4}}. A decrease in the disorder strength delays 
the divergence, but even a weak disorder amplitude ($R_d=0.01$) results in 
a sizable split (see~{\ref{fig4}}). The split occurs because the growth of 
initial fluctuations is drastically amplified by the presence of disorder 
({\ref{fig5}}).

Consider the growth of initial fluctuations, $H=1+\theta(\vec{X},T)$, 
in~\ref{tfe}. On Fourier-transforming $\theta(\vec{X},T)$ and integrating 
the resultant $\theta(\vec{k},T)$ over a uniform disorder distribution on 
$[R_m(1-R_d), R_m(1+R_d)]$, we obtain
\begin{equation}
 \bar{\theta}(\vec{k},T) = \exp\left[k^2\left\lbrace\alpha(R_m)-k^2\right\rbrace T\right]
 \frac{\sinh(bk^2R_mR_dT)}{bk^2R_mR_dT}\,\theta(\vec{k},0),
\label{theta}
\end{equation}
where $b=1-(1+D)^{-4}$. 
The factor which represents the effect of the disorder,
$\sinh(cT)/(cT)$, $\to 1$ as $R_d \to 0$, 
but amplifies the growth of initial fluctuations for nonzero $R_d$. In turn, 
this increases the sub-structure in the growing profile (see frames at 
$T=0.052, 10$ in~{\ref{fig5}}) -- resulting in a slower decrease of 
$N_d$ vs. $T$. 
The growing fluctuations are saturated in the late stages when domains 
are formed at the smaller value of the equilibrium thickness $H _m< 1$. 
Of course, the height field continues to grow for $H > 1$. These domains 
coarsen with time and obey the LS growth law 
\cite{pre10}. 
The presence of disorder causes local fluctuations in the domain 
structure -- see frame at $T=10000$ in~{\ref{fig5}}. 

In the late stages, the growth law is universal in~{\ref{fig4}} 
($N_d \sim T^{-1/3}$), regardless of the presence of disorder. This 
should be contrasted with the usual phase-separation problem, where 
coexisting domains are trapped by disordered sites 
\cite{pp92, ppr04, ppr05}. 
The subsequent coarsening proceeds by thermally-activated barrier 
hopping, leading to an asymptotically logarithmic growth law. In the 
SPS problem discussed here, the defects consist of ``interfaces'' 
between flat domains at $H=H_m$ and growing hills with $H \to \infty$. 
These defects do not become trapped as the disorder scale becomes 
irrelevant in comparison to the diverging height of the ``interfaces''. 
This late-stage universality has important implications for 
experimentalists. As mentioned earlier, it is not possible to eliminate 
disorder from experimental systems. However, our results show that 
disorder is irrelevant in the late stages, and all experiments will 
finally recover results for the pure case. The differences between 
the pure and disordered systems only show up in the early stages, 
as quantified above.

\begin{figure}[ht]
\begin{center}
\includegraphics*[width=0.8\textwidth]{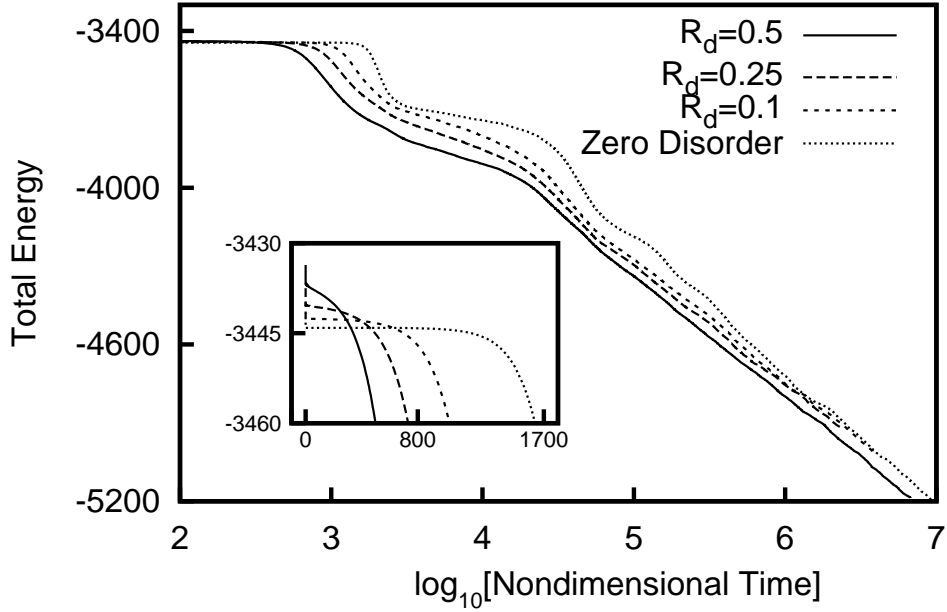}
\end{center}
\caption{Variation of the total free energy with nondimensional time for a
         system of size 2048 $\bar{L}_{M}$ and $D = 0.5$. 
         The curves present results for $R_m=-0.1$ and 
         $R_{d} =0.50,\,0. 25,\,0.1$. The other parameter values 
         are $P_{m}=\bar{L}_{M}/16$, $P_{d}=0.5$.
         The dotted curve corresponds to the case with zero disorder. 
         The inset shows the magnified view for energy variation in the early
         stages.
         }
\label{fig6}
\end{figure}

The evolution of the total free energy is also consistent  with the 
evolution of the defect density ({\ref{fig6}}). 
As expected, the overall free energy diminishes with time. The decay 
in the late stages is a universal power-law, which is independent of the 
disorder amplitude. The difference between the pure and disordered cases
is seen in the early stages, with faster decay for larger disorder values, 
as the growth of initial fluctuations is speeded up by disorder.

\begin{figure}[h]
\begin{center}
    \begin{tabular}{cc}
      \resizebox{80mm}{!}{\includegraphics{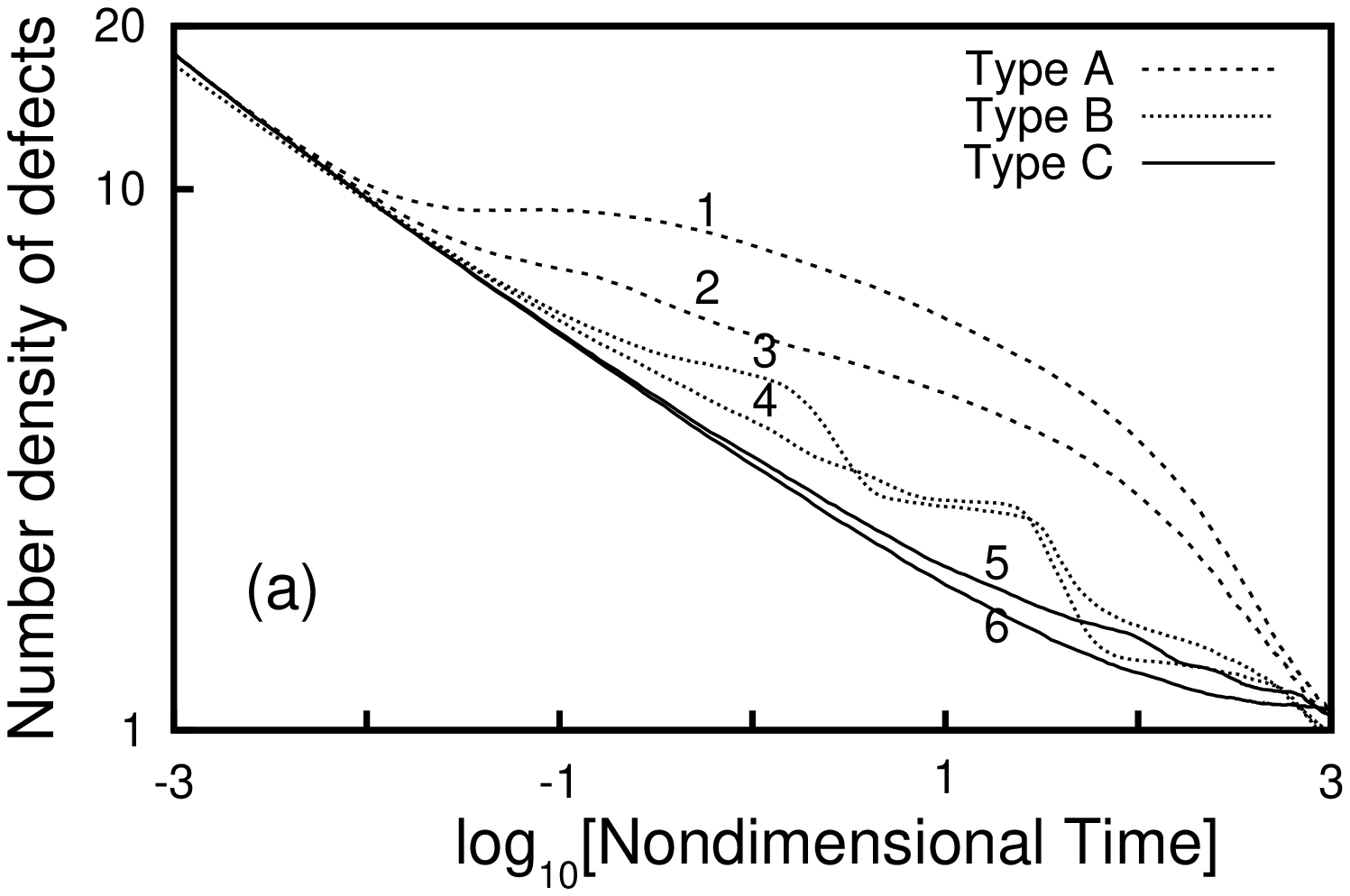}} & 
      \resizebox{80mm}{!}{\includegraphics{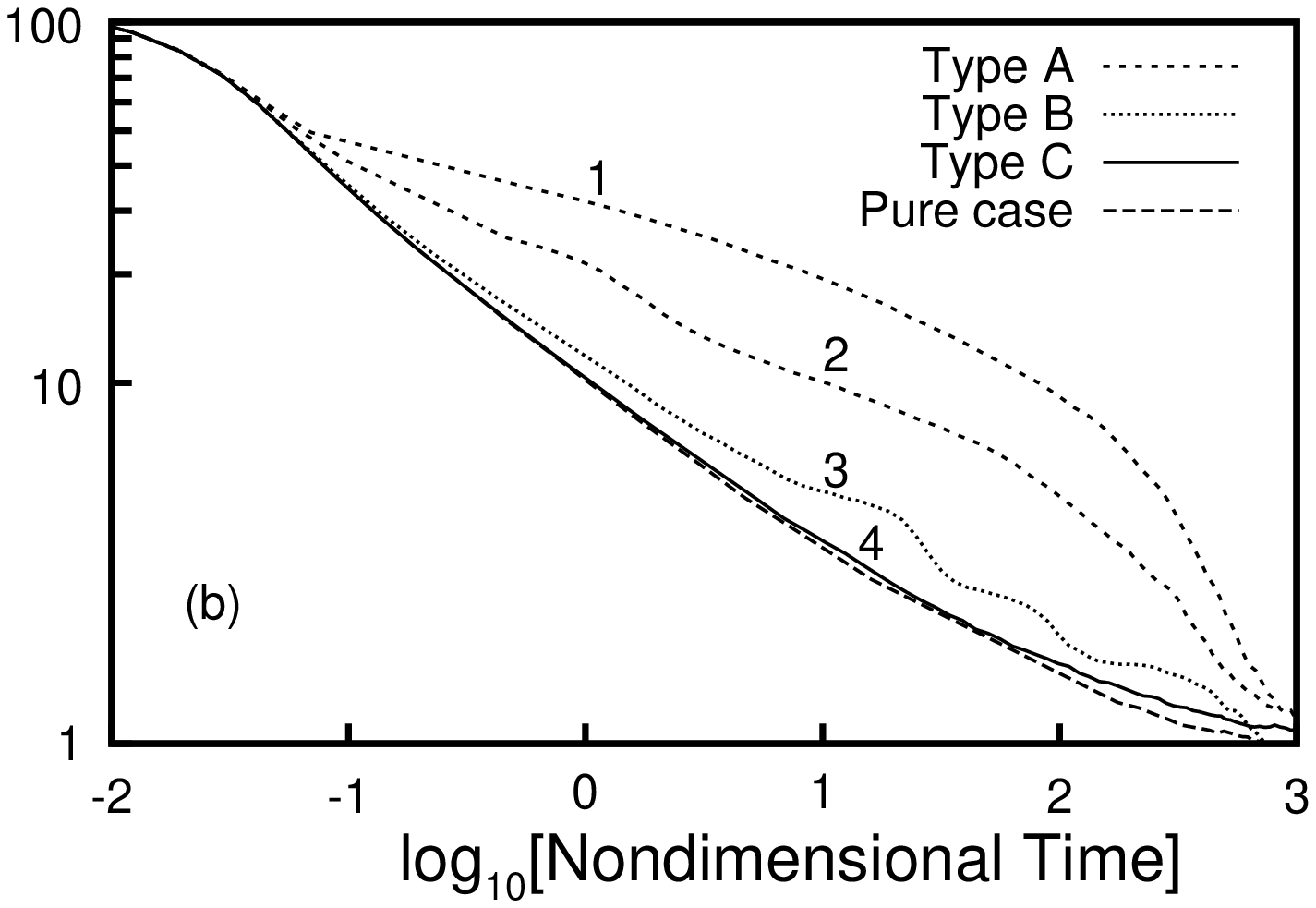}}
    \end{tabular}
\end{center}
\caption{Variation of the number density of defects with nondimensional time 
         in (a) $d=2$ and (b) $d = 3$. 
         The parameters are $R_{m}=-0.1$, $R_{d}=0.5$.
         The different curves are obtained by varying the mean patch
         size $P_m=f\bar{L}_M$. We classify the curves as follows: 
          Type~A ($f < 1/2$), Type~B ($1/2 < f < 2$), and 
          Type~C ($f > 2$). 
         }
\label{fig7}
\end{figure}

Next we focus on the effect of varying the patch size ($P_m, P_d$) on 
SPS. The early-stage kinetics exhibits three qualitatively 
distinct behaviors depending on the ratio $f~(P_m=f \bar{L}_M)$. 
Notice that the limit $P_m \to $ system size corresponds to the pure case. 
For small values of $f~(f < 1/2)$, the disorder leads to 
splitting as described earlier [Type~A curves in~{\ref{fig7}(a)}]. 
For intermediate values of $f~(1/2 < f < 2)$, the disorder leads 
to a distinctive staircase behavior for $N_d$ vs. $T$ [Type~B curves 
in~{\ref{fig7}(a)}]. In this case, the patch size is comparable to 
the spinodal wavelength, resulting in a stick-slip dynamics of the 
height field. For larger values of $f~(f > 2)$, the disorder 
has little effect and the early-stage kinetics resembles that for 
homogeneous substrates [Type~C curves in~{\ref{fig7}(a)}]. As 
expected, the late-stage dynamics is again universal with 
$N_d \sim T^{-1/3}$ -- we do not show this regime in~{\ref{fig7}(a)}.
A similar behavior is also seen in $d=3$ simulations --
in~{\ref{fig7}(b)}, 
we plot $N_d$ vs. $T$ for this case. The simulation details 
are provided in the figure caption.

As the spinodal length-scale can be controlled by changing the mean film
thickness ($l_{M} \propto h^{2}_{0}$, where $l_{M}$ is the dimensional
length-scale), one can use the present findings to address the inverse 
problem of assessing the disorder by thin-film experiments.
In thick films, we see Type~A behavior because $f \ll 1$. 
A reduction in the film thickness such that $1/2 < f < 2$ 
results in Type~B behavior.
A further reduction in thickness ($f > 2$) results in Type~C 
behavior.
We can use Type~C curves to first estimate $L_{M}$, and then
$R_{m}$ from~\ref{km}.
Thus, we can estimate the disorder amplitude to an order of magnitude.
An average value found by monitoring the spinodal length-scale when 
Type~B behavior starts and when it ends can fine-tune this estimate. 
The strength of the disorder can be found by any Type~C result.
%%%%%%%%%%%%%%%%%%%%%%%%%%%%%%%%%%%%%%%%%%%%%%%%%%%%%%%%%%%%%%%%%%%%%%%
\section{Conclusions}
In summary, we conclude that quenched chemical disorder has a pronounced 
effect on the early and intermediate stages of morphological phase 
separation in thin liquid films. However, the late-stage kinetics is 
universal and follows the Lifshitz-Slyozov (LS) growth law. These 
findings are in sharp contrast to the effects of quenched disorder seen 
in usual phase-separation processes, viz., the early stages remain 
undisturbed and domain growth is slowed down in the asymptotic regime. 
The early-stage kinetics shows different qualitative behavior, depending
on the relative sizes of disorder patches and the spinodal length-scale. 
Thus, the inverse problem of estimating disorder by thin-film 
experiments can also be addressed.

\bibliography{pccp_sub}

\end{document}